\documentclass[a4paper]{report}
\usepackage[utf8]{inputenc}
\usepackage[T1]{fontenc}
\usepackage{RJournal}
\usepackage{amsmath,amssymb,array}
\usepackage{booktabs}

\usepackage{amsfonts,amsthm,bbm,mathrsfs,mathrsfs,dsfont}
\usepackage{cancel,sectsty,color,enumitem,hyperref,graphicx}
\usepackage{ifthen}
\usepackage{subfig}

\newcommand{\bX}{\mathbf{X}}
\newcommand{\bv}{\mathbf{v}}
\newcommand{\bU}{\mathbf{U}}
\newcommand{\bx}{\mathbf{x}}
\newcommand{\bq}{\mathbf{q}}
\newcommand{\bp}{\mathbf{p}}
\newcommand{\bu}{\mathbf{u}}
\newcommand{\bbeta}{\boldsymbol{\beta}}
\newcommand{\bgamma}{\boldsymbol{\gamma}}
\newcommand{\bkappa}{\boldsymbol{\kappa}}
\newcommand{\bPi}{\boldsymbol{\Pi}}
\newcommand{\0}{\mathbf{0}}
\newcommand{\bSigma}{\boldsymbol{\Sigma}}

\defcitealias{Cattaneo-Farrell-Feng_2019_AoS}{CFF}

\begin{document}

\sectionhead{Contributed research article}
\volume{XX}
\volnumber{YY}
\year{20ZZ}
\month{AAAA}

\begin{article}
  \title{\pkg{lspartition}: Partitioning-Based Least Squares Regression}

\author{by Matias D. Cattaneo, 
    	Max H. Farrell and
	    Yingjie Feng
        }
    
\maketitle

\abstract{
Nonparametric partitioning-based least squares regression is an important tool in empirical work. Common examples include regressions based on splines, wavelets, and piecewise polynomials. 
This article discusses the main methodological and numerical features of the \textsf{R} software package \pkg{lspartition}, which implements modern estimation and inference results for partitioning-based least squares (series) regression estimation. This article discusses the main methodological and numerical features of the \textsf{R} software package \pkg{lspartition}, which implements results for partitioning-based least squares (series) regression estimation and inference from \citet*{Cattaneo-Farrell_2013_JoE} and \citet*{Cattaneo-Farrell-Feng_2019_AoS}. These results cover the multivariate regression function as well as its derivatives. First, the package provides data-driven methods to choose the number of partition knots optimally, according to integrated mean squared error, yielding optimal point estimation. Second, robust bias correction is implemented to combine this point estimator with valid inference. Third, the package provides estimates and inference for the unknown function both pointwise and \emph{uniformly} in the conditioning variables. In particular, valid confidence \emph{bands} are provided. Finally, an extension to two-sample analysis is developed, which can be used in treatment-control comparisons and related problems.}


\section{Introduction} \label{sec: intro}

Nonparametric partitioning-based least squares regression estimation is an important method for estimating conditional expectation functions in statistics, economics, and other disciplines. These methods first partition the support of covariates and then construct a set of local basis functions on top of the partition to approximate the unknown regression function or its derivatives. Empirically popular basis functions include splines, compactly supported wavelets, and piecewise polynomials. For textbook reviews on classical and modern nonparametric regression methodology see, among others, \citet*{Fan-Gijbels_1996_Book}, \citet*{Gyorfi-etal_2002_book}, \citet*{Ruppert-Wand-Carroll_2009_book}, and \citet*{Harezlak-Ruppert-Wand_2018_book}. For a review on partitioning-based approximations in nonparametrics and machine learning see \cite{Zhang-Singer_2010_Book} and references therein.

This article gives a detailed discussion of the software package \pkg{lspartition}, available for \textsf{R}, which implements partitioning-based least squares regression estimation and inference. This package offers several features which improve on existing tools, leveraging the recent results of \citet*{Cattaneo-Farrell_2013_JoE} and \citet*{Cattaneo-Farrell-Feng_2019_AoS}, and delivering data-driven methods to easily implement partitioning-based estimation and inference, including optimal tuning parameter choices and uniform inference results such as confidence bands. We cover splines, compactly supported wavelets, and piecewise polynomials, in a unified way, encompassing prior methods and routines previously unavailable without manual coding by researchers. Piecewise polynomials generally differ from splines and wavelets in that they do not enforce global smoothness over the partition, but in the special cases of zero-degree bases on a tensor-product partition, the three basis choices (i.e., zero-degree spline, Haar wavelet, and piecewise constant) are equivalent.

The first contribution offered by \pkg{lspartition} is a data-driven choice of the number of partitioning knots that is optimal in an integrated mean squared error (IMSE) sense. A major hurdle to practical implementation of any nonparametric estimator is tuning parameter choice, and by offering several feasible IMSE-optimal methods for splines, compactly supported wavelets, and piecewise polynomials, \pkg{lspartition} provides practitioners with tools to overcome this important implementation issue.

However, point estimation optimal tuning parameter choices yield invalid inference in general, and the IMSE-optimal choice is no exception. The second contribution of \pkg{lspartition} is the inclusion of robust bias correction methods, which allow for inference based on optimal point estimator. \pkg{lspartition} implements the three methods studied by \citet*{Cattaneo-Farrell-Feng_2019_AoS}, which are based on novel bias expansions therein. Both the bias and variance quantities are kept in pre-asymptotic form, yielding better bias correction and standard errors robust to conditional heteroskedasticity of unknown form. Collectively, this style of robust bias correction has been proven to yield improved inference in other nonparametric contexts \citep*{Calonico-Cattaneo-Farrell_2018_JASA,Calonico-Cattaneo-Farrell_2019_wp}.

The third main contribution is valid inference, both pointwise and uniformly in the support of the conditioning variables. When robust bias correction is employed, this inference is valid for the IMSE-optimal point estimator, allowing the researcher to combine an optimal partition for point estimation and a ``faithful'' measure of uncertainty (i.e., one that uses the same nonparametric estimation choices, here captured by the partition). In particular, \pkg{lspartition} delivers valid confidence \emph{bands} that cover the entire regression function and its derivatives. These data-driven confidence bands are constructed by approximating the distribution of $t$-statistic processes, using either a plug-in approach or a bootstrap approach. Importantly, the construction of confidence bands does not employ (asymptotic) extreme value theory, but instead uses the strong approximation results of \citet*{Cattaneo-Farrell-Feng_2019_AoS}, which perform substantially better in samples of moderate size.

Last but not least, the package also offers a convenient function to implement estimation and inference for linear combinations of regression estimators of different groups with all the features mentioned above. This function can be used to analyze conditional treatment effects in random control trials in particular, or for two-sample comparisons more generally. For example, a common question in applications is whether two groups have the same ``trend'' in a regression function, and this is often answered in a restricted way by testing a single interaction term in a (parametric) linear model. In contrast, \pkg{lspartition} delivers a valid measure of this difference nonparametrically and uniformly over the support of the conditioning variables, greatly increasing its flexibility in applications.

All of these contributions are fully implemented for splines, wavelets, and piecewise polynomials through the following four functions included in the package \pkg{lspartition}:
\begin{itemize}
	\item \code{lsprobust()}: This function implements estimation and inference for partitioning-based least squares regression. It takes the partitioning scheme as given, and constructs point and variance estimators, bias correction, conventional and robust bias-corrected confidence intervals, and simulation-based conventional and robust bias-corrected uniform inference measures (e.g., confidence bands). Three approximation bases are provided: B-splines, Cohen-Daubechies-Vial wavelets, and piecewise polynomials. When the partitioning scheme is not specified, the companion function \code{lspkselect()} is used to select a tensor-product partition in a fully data-driven fashion.
	
	\item \code{lspkselect()}: This function implements data-driven procedures to select the number of knots for partitioning-based least squares regression. It allows for evenly-spaced and quantile-spaced knot placements, and computes the corresponding IMSE-optimal choices. Two selectors are provided: rule of thumb (ROT) and direct plug-in (DPI) rule.
	
	\item \code{lsplincom()}: This function implements estimation and robust inference procedures for linear combinations of regression estimators of multiple groups based on \code{lsprobust()}. Given a user-specified linear combination, it offers all the estimation and inference methods available in the functions \code{lsprobust()} and \code{lspkselect()}.
	
	\item \code{lsprobust.plot()}: This function builds on \CRANpkg{ggplot2} \citep{ggplot2}, and is used as a wrapper for plotting results. It plots regression function curves, robust bias-corrected confidence intervals and uniform confidence bands, among other possibilities.
\end{itemize}

The paper continues as follows. Section \ref{sec: setup} describes the basic setup including a brief introduction to partitioning-based least squares regression and the empirical example to be used throughout to illustrate features of \pkg{lspartition}. Section \ref{sec: selection} discusses data-driven IMSE-optimal selection of the number of knots and gives implementation details. Estimation and inference implementation is covered in Section \ref{sec: estimation and inference}, including bias correction methods. The last section concludes. We defer to \citet*[][CFF hereafter]{Cattaneo-Farrell-Feng_2019_AoS} for complete theoretical and technical details. Statements below are sometimes specific versions of a general case therein.

\section{Setup} \label{sec: setup}

We assume that $\{(y_i, \bx_i')':1\leq i \leq n\}$ is an observed random sample of a scalar outcome $y_i$ and a $d$-vector of covariates $\bx_i\in\mathcal{X}\subset\mathbb{R}^d$. The object of interest is the regression function $\mu(\bx)=\mathbb{E}[y_i|\bx_i=\bx]$ or its derivative, the latter denoted by $\partial^{\bq}\mu(\bx)=\partial^{[\bq]}\mu(\bx)/\partial x_1^{q_1}\cdots\partial x_d^{q_d}$, for a $d$-tuple $\bq=(q_1, \ldots, q_d)'\in\mathbb{Z}_+^d$ with $[\bq]=\sum_{j=1}^{d}q_j$.

Estimation and inference is based on least squares regression of $y_i$ on set of basis functions of $\bx_i$ which are themselves built on top of a partition of the support $\mathcal{X}$. A partition, denoted by $\Delta=\{\delta_l\subset\mathcal{X}: 1\leq l \leq \kappa\}$, is a collection of $\kappa$ disjoint open sets such that the closure of their union is $\mathcal{X}$. For a partition, a set of basis functions, each of order $m$ and denoted by $\bp(\bx)$, is constructed so that each individual function (i.e., each element of the vector $\bp(\bx)$) is nonzero on a fixed number of contiguous $\delta_l$. \pkg{lspartition} allows for three such bases: piecewise polynomials, B-splines, and Cohen-Daubechies-Vial wavelets \citep*{Cohen_1993_ACHA}. For the first two bases, the order $m$ of the basis can be any positive integer, and any derivative of $\mu$ up to total order $(m-1)$ can be estimated employing such a basis. For wavelets, the current version allows for $m\leq 4$ (i.e., up to cubic wavelets), and $\bq=(0, \ldots, 0)$.  The package takes $m=2$ (linear basis) as default. To fix ideas, consider $d=1$ with piecewise polynomials. Each $\delta_l$ is an interval and $\bp(\bx)$ collects
all the indicator functions $\mathbf{1}\{x\in \delta_l\}$, $1\leq l\leq\kappa$.

Once the basis $\bp(\bx)$ is constructed, the final estimator of $\partial^\bq\mu(\bx)$ is
\begin{equation} \label{eq: point estimate}
\widehat{\partial^\bq\mu}(\bx)=\partial^\bq\bp(\bx)'\widehat{\bbeta}, \qquad \widehat{\bbeta}=\underset{\mathbf{b}\in\mathbb{R}^K}{\arg\min}\sum_{i=1}^{n}\left(y_i-\bp(\bx_i)'\mathbf{b}\right)^2,
\end{equation}
where $[\bq]<m$. When $\bq=\mathbf{0}$, we write $\widehat{\mu}(\cdot)=\widehat{\partial^\mathbf{0}\mu}(\cdot)$ for simplicity.

The approximation power of such estimators increases with the granularity of the partition $\Delta$ and the order $m$. We take the latter as fixed in practice. The most popular structure of $\Delta$ in applications is a tensor-product form, which partitions each covariate marginally into intervals and then sets $\Delta$ to be the set of all tensor (Cartesian) products of these intervals (\citetalias{Cattaneo-Farrell-Feng_2019_AoS} consider more general cases). For this type of partition, the user must choose the \emph{number} and \emph{placement} of the partitioning knots in each dimension. \pkg{lspartition} allows for three knot placement types: user-specified, evenly-spaced, and quantile-spaced. In the first case, the user has complete freedom to choose both the number and positions of knots for each dimension. In the latter two cases, the knot placement scheme is pre-specified, and hence only the number of subintervals for each dimension needs to be chosen.

We denote the number of knots in the $d$ dimensions of the regressor $\bx_i$ by $\bkappa=(\kappa_1, \ldots, \kappa_d)\in\mathbb{Z}_+^d$, which can be either specified by users or selected by data-driven procedures (see Section \ref{sec: selection} below). Moreover, for wavelet bases, motivated by the standard multi-resolution analysis, we provide an option \code{J} for the regression command \code{lsprobust()}, which indicates the resolution level of a wavelet basis. This gives $\kappa_\ell = 2^{J_\ell}, \ell = 1, \ldots d$, for a resolution $J_\ell$ \citep[see][for a review]{Chui_2016_Book}. In any case, the tuning parameter to be chosen is $\kappa = \kappa_1 \times \cdots \times \kappa_d$. In the next section we choose $\kappa$ to minimize the IMSE of the estimator \eqref{eq: point estimate}.

\subsection{Package and data}

We will showcase the main aspects of \pkg{lspartition} using a running empirical example. The package is available in \textsf{R} and can be installed as follows:

\begin{example*}
	\input{output/lspartition_1.txt}
\end{example*}
\vspace{-.1in}

The data we use come from Capital Bikeshare, and is available at \url{http://archive.ics.uci.edu/ml/datasets/Bike+Sharing+Dataset/}. For the first 19 days of each month of 2011 and 2012 we observe the outcome \code{count}, the total number of rentals and the covariates \code{atemp}, the ``feels-like'' temperature in Celsius, and \code{workingday}, a binary indicator for working days (versus weekends and holidays). The data is summarized as follows.

\begin{example*}
	\input{output/lspartition_2.txt}
\end{example*}
\vspace{-.1in}

We will investigate nonparametrically the relationship between temperature and number of rentals and compare the two groups defined by the type of days:

\begin{example*}
	\input{output/lspartition_3.txt}
\end{example*}
\vspace{-.2in}
The sample code that follows will use this designation of \code{y}, \code{x}, and \code{g}.

\section{Partitioning scheme selection} \label{sec: selection}

We will now briefly describe the IMSE expansion and its use in tuning parameter selection. To differentiate the original point estimator of \eqref{eq: point estimate} and the post-bias-correction estimators, we will add a subscript $``0"$ to the original estimator: $\widehat{\partial^\bq\mu}_0(\bx)$. The three bias corrections discussed below will add corresponding subscripts of 1, 2, and 3. We first discuss the bias and variance of $\widehat{\partial^\bq\mu}_0(\bx)$, and then use these for minimizing the IMSE. Throughout, $\approx$ denotes that the approximation holds for large sample in probability and $\mathbb{E}_n[\cdot]$ denotes the sample average over $1\leq i \leq n$. To simplify notation, we may write the estimator as \[\widehat{\partial^\bq\mu}_0(\bx) := \widehat{\bgamma}_{\bq,0}'\mathbb{E}_n[\bp(\bx_i) y_i], \quad\text{ where }\quad \widehat{\bgamma}_{\bq,0}(\bx)' := \partial^\bq \bp(\bx)'  \mathbb{E}_n[ \bp(\bx_i) \bp(\bx_i)']^{-1}.\]
Again, note the subscript ``0''; the bias-corrected estimators are of the same form (see below).

\subsection{Bias and variance}

The bias expansion for the $\widehat{\partial^\bq\mu}_0(\bx)$ is:
\begin{align}
\mathbb{E}[\widehat{\partial^\bq\mu}_0(\bx)|\bX] - \partial^{\bq}\mu(\bx)
&=\widehat{\bgamma}_{\bq,0}(\bx)'\mathbb{E}_n[\bp(\bx_i)\mu(\bx_i)] - \partial^{\bq}\mu(\bx) \label{eq: implicit bias}\\
&\approx
\mathscr{B}_{m, \bq}(\bx)-\widehat{\bgamma}_{\bq,0}(\bx)'\mathbb{E}_n[\bp(\bx_i)\mathscr{B}_{m, \0}(\bx_i)]. \label{eq: explicit bias}
\end{align}
$\mathscr{B}_{m, \bq}(\cdot)$ is the leading approximation error in the $L_\infty$-norm and the second term is the accompanying error from the linear projection of $\mathscr{B}_{m, \0}(\cdot)$ onto the space spanned by the basis functions. The form of each of these is complex, and depends on the basis, but what is crucial for the present purposes is that the form is known and the only unknown elements are derivatives of order $m$, $\partial^{\bu}\mu(\bx)$, $[\bu]=m$. \citetalias{Cattaneo-Farrell-Feng_2019_AoS} derive exact expressions for splines, wavelets, and piecewise polynomials. Both bias terms will, in general, contribute to the same order, and both will matter in finite samples. However, the second term in \eqref{eq: explicit bias} will be higher order if the bases are carefully constructed so that $\mathscr{B}_{m, \mathbf{0}}(\cdot)$ is orthogonal to $\bp(\cdot)$ in $L_2$ with respect to the Lebesgue measure. \pkg{lspartition} allows users to choose whether the projection of the leading error is used in partitioning scheme selection, as well as estimation and inference.

The conditional variance is straightforward from least squares algebra, and takes the familiar sandwich form:
\begin{equation}
\mathbb{V}[\widehat{\partial^{\bq}\mu}_0(\bx)|\bX]=\frac{1}{n}
\widehat{\bgamma}_{\bq,0}(\bx)'\bar{\bSigma}_0\widehat{\bgamma}_{\bq,0}(\bx), \qquad \bar{\bSigma}_0=\mathbb{E}_n\left[\bp(\bx_i)\bp(\bx_i)'\sigma^2(\bx_i)\right],
\end{equation}
where $\sigma^2(\bx_i)=\mathbb{V}\left[y_i|\bx_i\right]$. Only $\sigma^2(\bx_i)$ is unknown here, and will be replaced by a residual-based estimator. In particular \pkg{lspartition} allows for the standard Heteroskedasticity-Consistent (HC) class of estimators via the options \code{hc0}, \code{hc1}, \code{hc2}, \code{hc3}. See \citet{Long-Ervin_2000_AS} for a review in the context of least squares regression.

\subsection{Integrated mean squared error}

In general, for a weighting function $w(\bx)$, \citetalias{Cattaneo-Farrell-Feng_2019_AoS} derive the following (conditional) IMSE expansion:
\[\mathtt{IMSE}[\widehat{\partial^{\bq}\mu}(\cdot)|\bX] \approx
\frac{1}{n}\mathscr{V}_{\bkappa, \bq} +
\mathscr{B}_{\bkappa, \bq}, \qquad
\mathscr{V}_{\bkappa, \bq}  \asymp \kappa^{1+2[\bq]/d}, \qquad
\mathscr{B}_{\bkappa, \bq} \asymp \kappa^{2(m-[\bq])/d},
\]
where the $n$-varying quantities $\mathscr{V}_{\bkappa, \bq}$ and $\mathscr{B}_{\bkappa, \bq}$ correspond to a fixed-$n$ approximation to the variance and squared bias, respectively.

Under regularity conditions on the partition and basis used, \citetalias{Cattaneo-Farrell-Feng_2019_AoS} derive explicit leading constants in this expansion. \pkg{lspartition} implements IMSE-minimization for the common simple case where $\Delta$ is a tensor product of marginally formed intervals where the same number of intervals are used for each dimension. Specifically, $\Delta_\ell=\{\underline{x}_\ell = t_{\ell, 0} < t_{\ell,1} < \cdots <t_{\ell, \bar{\kappa}-1} < t_{\ell, \bar{\kappa}} = \bar{x}_\ell\}$ partitions $\mathcal{X}_\ell$ into $\bar{\kappa}$ subintervals, and the complete partition $\Delta=\otimes_{\ell=1}^d\Delta_\ell$ where $\otimes$ denotes tensor (Cartesian) product. Thus, the IMSE-optimal number of cells of a tensor-product partition is $\kappa_{\mathtt{IMSE}}= \bar{\kappa}_{\mathtt{IMSE}}^d \asymp n^{\frac{d}{2m+d}}$.

To select $\bar{\kappa}_{\mathtt{IMSE}}$, or equivalently $\kappa_{\mathtt{IMSE}}$, assume that the partitioning knots $\{0=t_{\ell, 0}<t_{\ell,1}<\cdots<t_{\ell, \bar{\kappa}-1} < t_{\ell, \bar{\kappa}}=1\}$ are generated as quantiles of some marginal distributions $G_\ell(\cdot)$,
$\ell=1, \ldots, d$, that is,
\[t_{\ell, l} = G_\ell^{-1}\left(\frac{l}{\bar{\kappa}}\right), \quad l=0, 1, \ldots, \bar{\kappa}, 
\quad \ell=1, \ldots, d.
\]
where $G_\ell^{-1}(v) = \inf\{x\in\mathbb{R}: G_\ell(x)\geq v \}$. Then, the IMSE-optimal choice for $\bq=\mathbf{0}$ is
\[\bar{\kappa}_{\mathtt{IMSE},\mathbf{0}}=\bigg\lceil \left(\frac{2m \mathscr{B}_{G,\0}}{d\mathscr{V}_{\0}}\right)^{\frac{1}{2m+d}}n^{\frac{1}{2m+d}}\bigg\rceil
\]
where $\lceil x \rceil$ is a ceiling operator that outputs the smallest integer that is no less than $x$ and $\mathscr{B}_{G,\0}$ is a (squared) bias term that may depend on the marginals $G_\ell$ and, as before, is entirely known up to $m^{th}$ order derivatives: $\partial^{\bu}\mu(\bx)$, $[\bu]=m$.

\subsection{Implementation details}

Two popular choices of partitioning schemes are evenly-spaced partitions (\code{ktype="uni"}), which sets $G_\ell(\cdot)$ to be the uniform distribution over the support of the data, and quantile-spaced partitions (\code{ktype="qua"}), which sets $G_\ell(\cdot)$ to be the empirical distribution function of each covariate. The package \pkg{lspartition} implements both partitioning schemes, and for each case offers two IMSE-optimal tuning parameter selection procedures: rule of thumb (\code{imse-rot}) and direct plug-in (\code{imse-dpi}) choices. We close this section with a brief description of the implementation details and an illustration using real data.

\medskip
\noindent\strong{Rule-of-Thumb Choice}
\smallskip

The rule-of-thumb choice is based on the special case of $\bq=\0$. Let the weighting function $w(\bx)$ be the density of $\bx_i$. The implementation steps are summarized in the following:

\begin{itemize}
	\item \textbf{Bias constant}.
	The unknown quantities in the bias constants are: $\partial^{\bu}\mu(\cdot)$, $\bu \in \Lambda_m$, which is estimated by a global polynomial regression of degree $(m+2)$; and the density of $\bx_i$, which is either assumed to be uniform or estimated by a trimmed-from-below Gaussian reference model (controlled by the option \code{rotnorm}).

	\item \textbf{Variance constant}.
	The unknown quantities in the variance constants are: the conditional variance
	$\sigma^2(\bx)=\mathbb{E}[y_i^2|\bx_i=\bx]-(\mathbb{E}[y_i|\bx_i=\bx])^2$, which is estimated by global polynomial regressions of degree $(m+2)$; and the density of $\bx_i$, which is either assumed to be uniform or estimated by a trimmed-from-below Gaussian reference model.	

	\item \textbf{Rule-of-thumb $\hat{\bar{\kappa}}_{\mathtt{rot}}$}.
	Using the above results, a simple rule-of-thumb choice of $\bar\kappa$ is
	\[
	\hat{\bar\kappa}_{\mathtt{rot}}=\bigg\lceil\left(
	\frac{2m \widehat{\mathscr{B}}_{G,\0} }
	{d\widehat{\mathscr{V}}_{\mathbf{0}}}
	\right)^{\frac{1}{2m+d}} n^{\frac{1}{2m+d}}\bigg\rceil.
	\]
	where $\widehat{\mathscr{B}}_{G,\0}$ and $\widehat{\mathscr{V}}_{\mathbf{0}}$ are the estimates of bias and variance constants respectively.
	While this choice of $\bar\kappa$ is obtained under strong parametric assumptions, it still exhibits the correct convergence rate ($\hat{\bar\kappa}_{\mathtt{rot}}\asymp n^{\frac{1}{2m+d}}$).
\end{itemize}

The command \code{lspkselect()} implements the rule-of-thumb selection (\code{kselect="imse-rot"}). For example, we focus on a subsample of bike rentals during working days (\code{g==1}), and then the selected number of knots are reported in the following:

\begin{example*}
	\input{output/lspartition_4.txt}
\end{example*}
\vspace{-.2in}

In this example, for the point estimator based on an evenly-spaced partition, the rule-of-thumb estimate of the IMSE-optimal number of knots is $\mathtt{k}=5$, and for the derivative estimators used in bias correction for later inference, the rule-of-thumb choice is $\mathtt{k.bc=9}$.
\medskip

\noindent\strong{Direct Plug-in Choice}
\smallskip

Assuming that the weighting $w(\bx)$ is equal to the density of $\bx_i$, the package \pkg{lspartition} implements a direct-plug-in (DPI) procedure summarized by the following steps.
\begin{itemize}
	\item \textbf{Preliminary choice of $\bar\kappa$}:
	Implement the rule-of-thumb procedure to obtain $\hat{\bar\kappa}_{\mathtt{rot}}$.
	
	\item \textbf{Preliminary regression}.
	Given the user-specified basis, knot placement scheme, and rule-of-thumb choice $\hat{\bar\kappa}_{\mathtt{rot}}$, implement a partitioning-based regression of order $(m+1)$ to estimate all necessary order-$m$ derivatives; denote these by $\widehat{\partial^\bu\mu}_{\mathtt{pre}}(\cdot)$, $[\bu]=m$.
	
	\item \textbf{Bias constant}.
	Construct an estimate $\widehat{\mathscr{B}}_{m,\bq}(\cdot)$ of the leading error $\mathscr{B}_{m,\bq}(\cdot)$ by replacing $\partial^\bu\mu(\cdot)$ by $\widehat{\partial^\bu\mu}_{\mathtt{pre}}(\cdot)$.
	$\widehat{\mathscr{B}}_{m, \mathbf{0}}(\cdot)$ can be obtained similarly. Then, use the pre-asymptotic version of the conditional bias to estimate the bias constant:
	\[\widehat{\mathscr{B}}_{\bkappa,\bq} =
	\frac{1}{n}\sum_{i=1}^{n}\left(\widehat{\mathscr{B}}_{m, \bq}(\bx_i)
	-\widehat{\bgamma}_{\bq,0}(\bx_i)'\mathbb{E}_n[\bp(\bx_i)\mathscr{B}_{m,\mathbf{0}}(\bx_i)]\right)^2.
	\]
	As mentioned before, for the three bases considered in the package \pkg{lspartition}, the second term in the conditional bias is of smaller order under some additional conditions. We employ this property to simplify the estimate of bias constant for wavelets. For splines and piecewise polynomials, however, users may specify whether the projection of the leading error is taken into account in the selection procedure (see option \code{proj}).
	
	\item \textbf{Variance constant}.
	Implement a partitioning-based series regression of order $m$ with $\bar\kappa=\hat{\bar\kappa}_{\mathtt{rot}}$, and then use the pre-asymptotic version of the conditional variance to estimate the variance constant. Specifically, let
	\begin{equation*}
	\widehat{\mathscr{V}}_{\bkappa, \bq} =
	\frac{1}{n}\sum_{i=1}^{n}\widehat{\bgamma}_{\bq,0}(\bx_i)'\widehat{\bSigma}_0
	\widehat{\bgamma}_{\bq,0}(\bx_i), \quad\text{and}\quad
	\widehat{\bSigma}_0=\mathbb{E}_n[\bp(\bx_i)\bp(\bx_i)' w_i \widehat{\epsilon}^2_i]
	\end{equation*}
	where $\widehat{\epsilon}_i$'s are regression residuals, $\widehat{\bSigma}_0$ is an estimate of $\bSigma_0=\mathbb{E}[\bp(\bx_i)\bp(\bx_i)'\sigma^2(\bx_i)]$, and $w_i$ is the weighting scheme used to construct different HC variance estimators.
	
	\item \textbf{Direct plug-in $\bar\kappa$}.
	Collecting all these results, a direct plug-in choice of $\bar\kappa$ is
	\[\hat{\bar\kappa}_{\mathtt{dpi}}=
	\bigg\lceil\left(\frac{2(m-[\bq])\hat{\bar\kappa}_{\mathtt{rot}}^{2(m-[\bq])}\widehat{\mathscr{B}}_{\bkappa,\bq}}{(d+2[\bq])\hat{\bar\kappa}_{\mathtt{rot}}^{-(d+2[\bq])}\widehat{\mathscr{V}}_{\bkappa,\bq}}\right)^{\frac{1}{2m+d}} n^{\frac{1}{2m+d}}\bigg\rceil.
	\]
\end{itemize}

The following shows the results of the direct plug-in procedure based on the real data:

\begin{example*}
	\input{output/lspartition_5.txt}
\end{example*}
\vspace{-.1in}

The direct plug-in procedure gives more partitioning knots than the rule-of-thumb, leading to a finer partition. For point estimation, $\hat{\bar\kappa}_{\mathtt{dpi}}=8$ knots are suggested, while for bias correction purpose, it selects $\hat{\bar\kappa}_{\mathtt{dpi}}=10$ knots to estimate derivatives in the leading bias. Quantile-spaced knot placement is obtained by adding \code{ktype = "qua"}.
%

\section{Estimation and inference} \label{sec: estimation and inference}

This section reviews and illustrates the estimation and inference procedures implemented. A crucial ingredient is the bias correction that allows for valid inference after tuning parameter selection.

\subsection{Point estimation and bias correction}

The estimator $\widehat{\partial^{\bq}\mu}_0(\bx)$ is IMSE-optimal from a point estimation perspective when implemented using the choice $\kappa_{\mathtt{IMSE}}$ to form $\Delta$, but conventional inference methods based on this resulting point estimator will be invalid. More precisely, the ratio of bias to standard error in the $t$-statistic is non-negligible, requiring either ad-hoc undersmoothing or some form of bias correction. In addition to the (uncorrected) point estimate in \eqref{eq: point estimate}, the package \pkg{lspartition} implements the three bias correction options derived by \citetalias{Cattaneo-Farrell-Feng_2019_AoS} for valid (pointwise and uniform) inference. All these strategies resort to a higher-order basis, $\tilde{\bp}(\bx)$, of order $\tilde{m}>m$. The partition $\tilde{\Delta}$ where $\tilde{\bp}(\bx)$ is built on may be different from $\Delta$ but need not be. These approaches allow researchers to combine an optimal point estimate $\widehat{\partial^\bq \mu}_0(\bx)$ based on the IMSE-optimal $\kappa_{\mathtt{IMSE}}$ with inference based on the same tuning parameter and partitioning scheme choices.

Our bias correction strategies are based on \eqref{eq: implicit bias} and \eqref{eq: explicit bias}, where the only unknowns are $\mu(\cdot)$, $\partial^{\bq}\mu(\cdot)$, and $\partial^{\bu}\mu(\cdot)$ for $[\bu]=m$. These are summarized as follows; see \citetalias{Cattaneo-Farrell-Feng_2019_AoS} for details.
\begin{itemize}
	\item \textbf{Approach 1: Higher-order-basis bias correction}. Use $\tilde{\bp}(\bx)$ to construct a higher-order least squares estimator $\widehat{\partial^\bq\mu}_1(\bx)$ which takes  exactly the same form as
	$\widehat{\partial^\bq\mu}_0(\bx)$ but has less bias. If we substitute $y_i$ and $\widehat{\partial^{\bq}\mu}_1(\bx)$ for $\mu(\bx_i)$ and $\partial^\bq\mu(\bx)$ in \eqref{eq: implicit bias} respectively and  subtract this estimated bias from $\widehat{\partial^\bq\mu}_0(\bx)$, the resulting ``bias-corrected'' estimator is equivalent to $\widehat{\partial^{\bq}\mu}_1(\bx)$. This option is called by \code{bc="bc1"}.
	
	\item \textbf{Approach 2: Least squares bias correction}. Construct $\widehat{\partial^{\bq}\mu}_1(\bx)$ and substitute it for $\partial^\bq\mu(\bx)$ in \eqref{eq: implicit bias}, but replace $\mu(\bx_i)$ by $\hat{\mu}_1(\bx_i)$ rather than $y_i$. The least squares bias-corrected estimator $\widehat{\partial^\bq\mu}_2(\bx)$ is obtained by subtracting this estimated bias from $\widehat{\partial^\bq\mu}_0(\bx)$. The supplement to \citetalias{Cattaneo-Farrell-Feng_2019_AoS} discusses in detail how this approach relates to higher-order-basis bias correction and when they are equivalent. This option is called by \code{bc="bc2"}.
	
	\item \textbf{Approach 3: Plug-in bias correction}. Referring to \eqref{eq: explicit bias}, use $\tilde{\bp}(\bx)$ to construct $\widehat{\partial^{\bu}\mu}_1(\bx)$ for all needed $\bu$.  Substitute $\widehat{\partial^{\bu}\mu}_1(\bx)$ and $\widehat{\partial^{\bu}\mu}_1(\bx_i)$ for $\partial^{\bu}\mu(\bx)$ and $\partial^{\bu}\mu(\bx_i)$ in $\mathscr{B}_{m, \bq}(\bx)$ and $\mathscr{B}_{m, \0}(\bx_i)$ respectively. Subtracting this estimated bias from $\widehat{\partial^\bq\mu}_0(\bx)$ leads to a plug-in bias-corrected estimator $\widehat{\partial^\bq\mu}_3(\bx)$. This option is called by \code{bc="bc3"}.
\end{itemize}
The optimal (uncorrected) point estimator ($j=0$) and the three bias-corrected estimators ($j=1,2,3$) can be written in a unified form:
\begin{equation*}
\widehat{\partial^\bq\mu}_j(\bx)=\widehat{\bgamma}_{\bq,j}(\bx)'\mathbb{E}_n[\bPi_j(\bx_i)y_i], \quad j=0,1,2,3.
\end{equation*}
These estimators only differ in $\widehat{\bgamma}_{\bq,j}(\cdot)$ and $\bPi_j(\cdot)$, which depend in different ways on $\bp(\bx)$ and $\tilde{\bp}(\bx)$. See \citetalias{Cattaneo-Farrell-Feng_2019_AoS} for exact formulas.

\subsection{Pointwise inference}

Pointwise inference relies on a Gaussian approximation for the $t$-statistics:
\[\widehat{T}_j(\bx)=
\frac{\widehat{\partial^{\bq}\mu}_j(\bx)-\partial^{\bq}\mu(\bx)}
{\sqrt{\widehat{\Omega}_j(\bx)/n}}
\rightsquigarrow \mathsf{N}(0, 1),
\quad j=0,1,2,3,
\]
where $\widehat{\Omega}_j(\bx)/n=\widehat{\bgamma}_{\bq,j}(\bx)'\widehat{\bSigma}_j\widehat{\bgamma}_{\bq,j}(\bx)/n$ is an estimator of the conditional variance of $\widehat{\partial^\bq\mu}_j(\cdot)$, and $\rightsquigarrow$ denotes convergence in distribution. $\widehat{\bSigma}_j(\bx)=\mathbb{E}_n[\bPi_j(\bx_i)\bPi_j(\bx_i)'w_i\widehat{\epsilon}_{i,j}^2]$ is a consistent estimator of $\bSigma_j=\mathbb{E}[\bPi_j(\bx_i)\bPi_j(\bx_i)\sigma^2(\bx_i)]$ where $\widehat{\epsilon}_{i,j}=y_i-\widehat{\mu}_j(\bx_i)$ and the $w_i$'s are additional weights leading to various HC variance estimators.
Then nominal $100(1-\alpha)$-percent symmetric confidence intervals are
\begin{equation} \label{eq: CI}
I_j(\bx)=\left[\widehat{\partial^\bq\mu}_j(\bx)-\Phi_{1-\alpha/2}
\sqrt{\widehat{\Omega}_j(\bx)/n},\quad \widehat{\partial^\bq\mu}_j(\bx)-
\Phi_{\alpha/2}\sqrt{\widehat{\Omega}_j(\bx)/n}\right]
\end{equation}
where $\Phi_u$ is the $u^{th}$ quantile of the standard normal distribution.

For conventional confidence intervals ($j=0$), the (asymptotically) correct coverage relies on undersmoothing ($\kappa \gg \kappa_{\mathtt{IMSE}}$) that renders the bias negligible relative to the standard error in large samples. Though straightforward in theory, it is difficult to implement in a principled way. In comparison, given the IMSE-optimal tuning parameter, all three bias-corrected estimators ($j=1,2,3$) have only higher-order bias, and thus the corresponding confidence intervals based on these estimators will have asymptotically correct coverage. Importantly, the Studentization quantity $\widehat{\Omega}_j(\bx)/n$ also captures the additional variability introduced by bias correction.

We now illustrate the pointwise inference features of \code{lsprobust()} using the bike rental data. The previous result of knot selection based on the DPI procedure will be employed. Specifically, we set \code{nknot=8} for point estimation. For higher-order-basis bias correction (\code{bc="bc1"}), the same number of knots is used to correct bias by default, while for plug-in bias correction (\code{bc="bc3"}), we use $10$ knots (\code{bnknot=10}) to estimate the higher-order derivatives in the leading bias. One may leave these options unspecified and then the command \code{lsprobust()} will automatically implement knot selection using the command \code{lspkselect()}.

\begin{example*}
\input{output/lspartition_7.txt}
\end{example*}
\vspace{-.1in}

The above table summarizes the results for pointwise estimation and inference, including point estimates, conventional standard errors, and robust confidence intervals based on higher-order-basis bias correction for $20$ quantile-spaced evaluation points. We can use the companion plotting command \code{lsprobust.plot()} to visualize the results:

\begin{example*}
\input{output/lspartition_8.txt}
\end{example*}
\vspace{-.2in}
\begin{figure}[!ht]
	\centering
	\subfloat[Higher-order-basis bias correction \label{figure: ci-bc1}]{\includegraphics[width=0.47\textwidth]{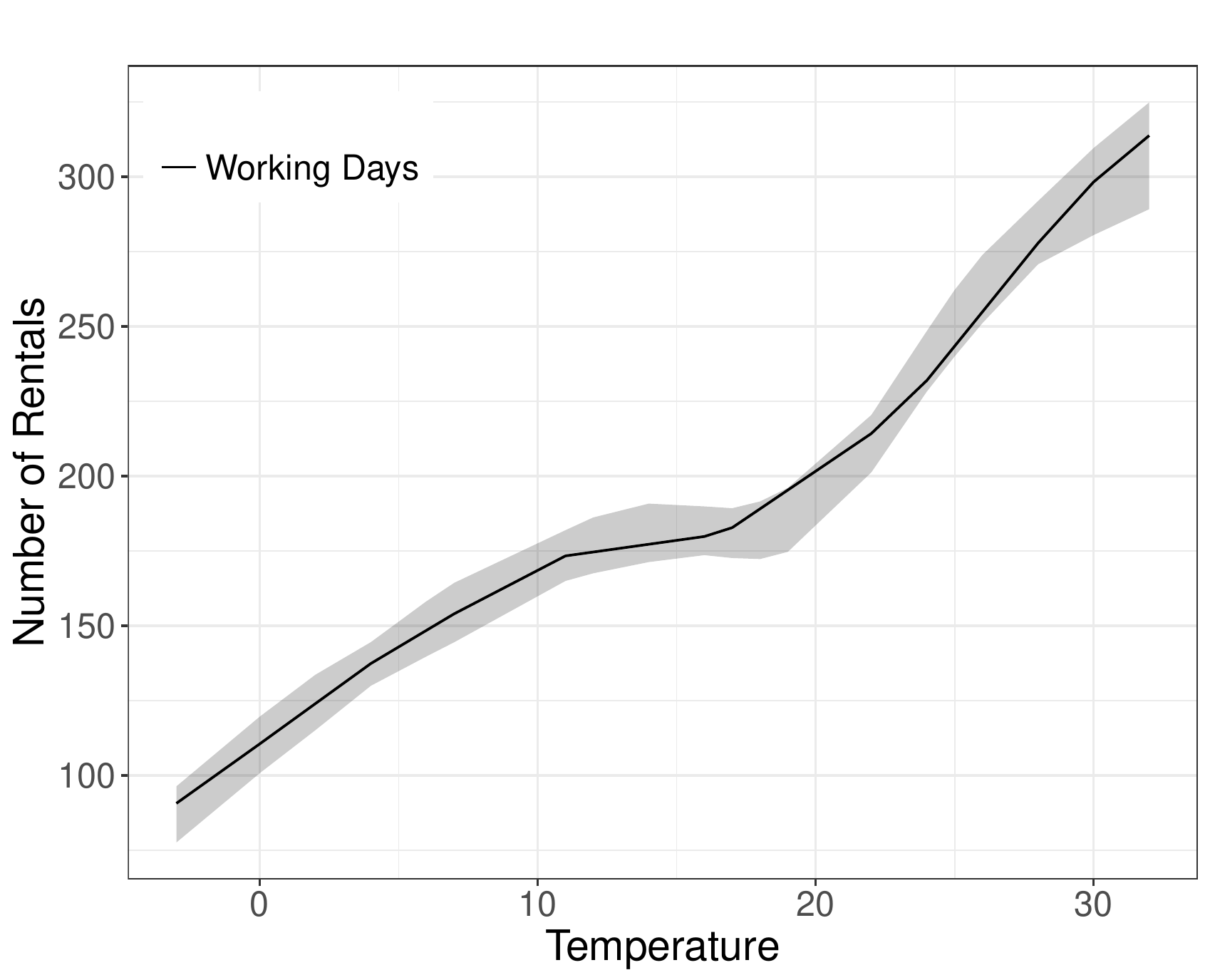}}
	~
	\subfloat[Plug-in bias correction \label{figure: ci-bc3}]{\includegraphics[width=0.47\textwidth]{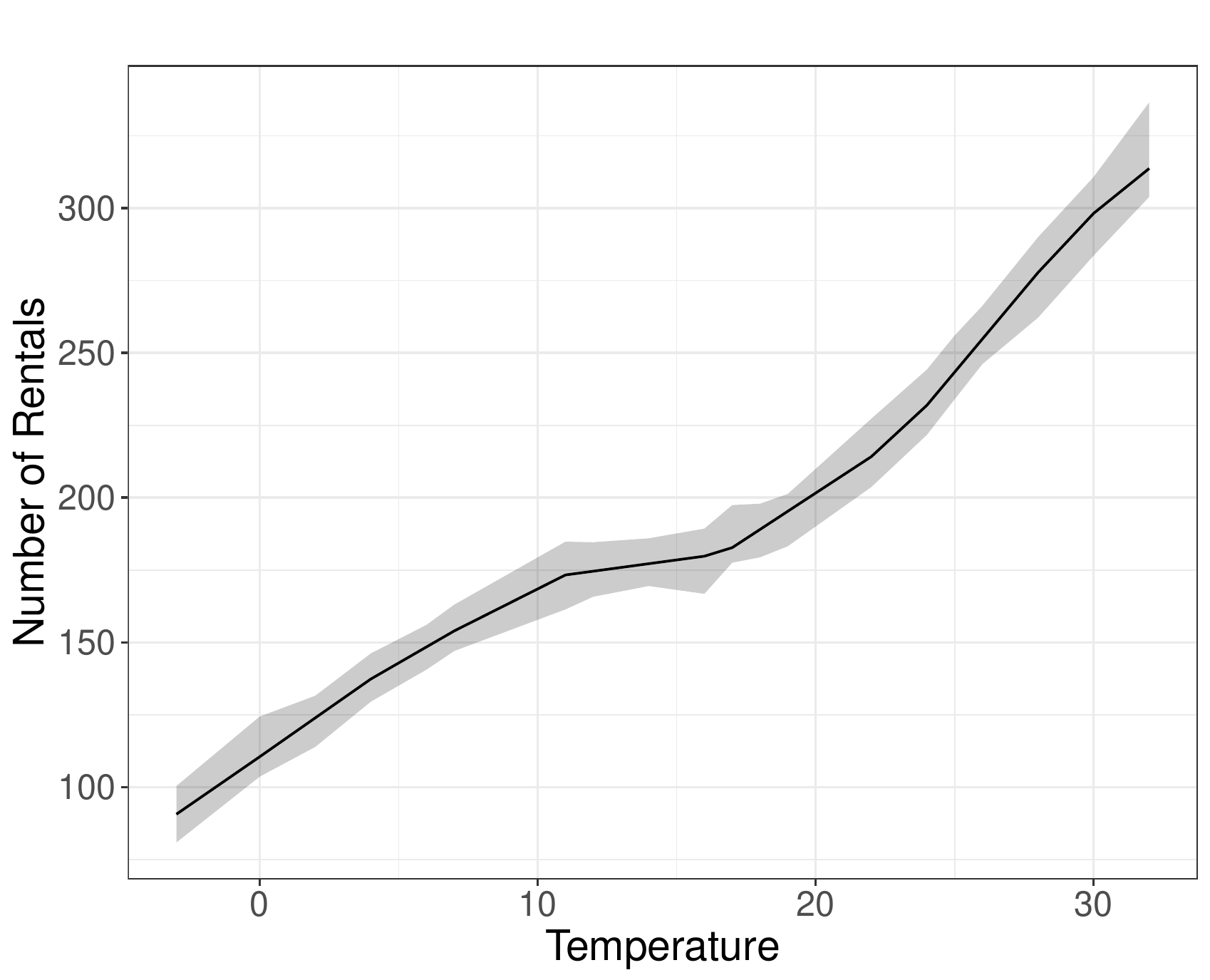}}
	\caption{Point estimation and pointwise robust confidence intervals}
	\label{figure: pointwise}
\end{figure}

The result is displayed in Figure \ref{figure: pointwise}. As the temperature gets higher, the number of rentals increases as expected. Both panels show the same point estimator, $\widehat{\mu}_0$. We plot both the robust confidence intervals based on higher-order-basis bias correction (Figure \ref{figure: ci-bc1}) and plug-in bias correction (Figure \ref{figure: ci-bc3}). Since the higher-order-basis approach is equivalent to a quadratic spline fitting, the resulting confidence interval has a smoother shape.

\subsection{Uniform inference}

To obtain uniform inference (over the support of $\bx$), \citetalias{Cattaneo-Farrell-Feng_2019_AoS} establish Gaussian approximations for the \emph{whole} $t$-statistic processes, and propose several sampling-based approximations which are easy to implement in practice. To be concrete, for each $j=0,1,2,3$, there exists a Gaussian process $Z_j(\cdot)$ such that
\[\widehat{T}_j(\cdot)\approx_d Z_j(\cdot), \quad Z_j(\cdot)=\frac{\bgamma_{\bq,j}(\cdot)'\bSigma_j^{1/2}}{\sqrt{\Omega_j(\cdot)}}\mathsf{N}_{K_j}, \quad K_j=\dim(\bPi_j(\cdot))
\]
where $\bgamma_{\bq,j}(\cdot)$ and $\Omega_j(\cdot)$ are population counterparts of $\widehat{\bgamma}_{\bq,j}(\cdot)$ and $\widehat{\Omega}_j(\cdot)$, $\mathsf{N}_{K_j}$ is a $K_j$-dimensional standard normal random vector, and $K_j$ is the length of $\bPi_j$, and is proportional to $\kappa$. The notation $\approx_d$ means that the two processes are asymptotically equal in distribution in the following sense: in a sufficiently rich probability space, we have identical copies of $\widehat{T}_j(\cdot)$ and $Z_j(\cdot)$ whose difference converges in probability to zero uniformly.

The Gaussian stochastic process $Z_j(\cdot)$ is not feasible in practice because it involves unknown population quantities. Thus, the package \pkg{lspartition} offers two options for implementation: \emph{plug-in} or \emph{bootstrap}.
\begin{itemize}
	\item \textbf{Plug-in}. Replace all unknowns in $Z_j(\cdot)$ by some consistent estimators:
	\[\widehat{Z}_j(\cdot)=
	\frac{\widehat{\bgamma}_{\bq,j}(\cdot)'\widehat{\bSigma}_j^{1/2}}
	{\sqrt{\widehat{\Omega}_j(\cdot)}} \mathsf{N}_{K_j}.\]
	\citetalias{Cattaneo-Farrell-Feng_2019_AoS} show that $\widehat{Z}_j(\cdot)$ delivers a valid distributional approximation to $\widehat{T}_j(\cdot)$. In practice one may obtain many simulated realizations of $\widehat{Z}_j(\cdot)$ by sampling from the $K_j$-dimensional standard normal distribution \emph{conditional on the data}. This option is called by \code{uni.method="pl"}.
	
	\item \textbf{Bootstrap}. Construct a bootstrapped version of the approximation process (conditional on the data):
	\[\widehat{z}_j^*(\cdot)= \frac{\widehat{\bgamma}_{\bq,j}(\cdot)'\mathbb{E}_n[\bPi_j(\bx_i)\widehat{\epsilon}_{i,j}^*]}
	{\sqrt{\widehat{\Omega}^*_j(\cdot)/n}}
	\]
	where $\widehat{\Omega}^*_j(\cdot)=\widehat{\bgamma}_{\bq,j}(\cdot)'\mathbb{E}_n[\bPi_j(\bx_i)\bPi_j(\bx_i)'(\widehat{\epsilon}^*_{i,j})^2]\widehat{\bgamma}_{\bq,j}(\cdot)$, $\widehat{\epsilon}_{i,j}^*=\omega_i\widehat{\epsilon}_{i,j}$ and $\{\omega_i\}_{i=1}^n$ is an i.i.d sequence of bounded random variables with zero mean and unit variance. \citetalias{Cattaneo-Farrell-Feng_2019_AoS} show that this bootstrapped process also approximates $Z_j(\cdot)$ conditional on the data. Thus one can implement bootstrapping by sampling from the distribution of $\omega_i$ given the data. In the package \pkg{lspartition}, the $\omega_i$'s are taken to be Rademacher variables, and this option is called by  \code{uni.method="wb"}.
\end{itemize}

Importantly, these strong approximations apply to the whole $t$-statistic processes, and thus can be used to implement general inference procedures based on transformations of $\widehat{T}_j(\cdot)$. The main regression command \code{lsprobust()} will output the the following quantities for uniform analyses upon setting \code{uni.out=TRUE}:
\begin{itemize}
	\item \code{t.num.pl, t.num.wb1, t.num.wb2}: The numerators of approximation processes except the ``simulated components'', which are evaluated at a set of pre-specified grid points $\mathcal{K}$. Suppose that $\mathcal{K}$ contains $L$ grid points. Then for the plug-in method, the numerator, stored in \code{t.num.pl}, is the $L\times K_j$ matrix   $\left\{\widehat{\bgamma}_{\bq,j}(\bx)'\widehat{\bSigma}_j^{1/2}/\sqrt{n}:
	\bx\in\mathcal{K}\right\}$.
	For wild bootstrap, the numerator is separated to  \code{t.num.wb1} and \code{t.num.wb2}, which are  $\left\{\widehat{\bgamma}_{\bq,j}(\bx)'/n: \bx\in\mathcal{K}\right\}$  and $(\bPi_j(\bx_1),\ldots, \bPi(\bx_n))'$ respectively.
	
	\item \code{t.denom}: The denominator of approximation processes, i.e., $\Big\{\sqrt{\widehat{\Omega}_j(\bx)/n}:\bx\in\mathcal{K}\Big\}$, stored in a vector of length $L$.
	
	\item \code{res}: Residuals from the specified bias-corrected regression (needed for bootstrap-based approximation).
\end{itemize} 

For example, the following command requests the necessary quantities for uniform inference based on the plug-in method:

\begin{example*}
\input{output/lspartition_9.txt}
\end{example*}
\vspace{-.1in}

We list the first $5$ rows of the numerator matrix. Each row corresponds to a grid point. Since we use a linear spline for point estimation and set \code{nknot=4}, the higher-order-basis bias correction is equivalent to quadratic spline fitting. Thus the numerator matrix has $7$ columns corresponding to the quadratic spline basis.

As a special application, these results can be used to construct uniform confidence bands, which builds on the suprema of $|\widehat{T}_j(\cdot)|$. The function \code{lsprobust()} computes the critical value to construct confidence bands. Specifically, it generates many simulated realizations of $\widehat{Z}_j(\cdot)$ or $\widehat{z}^*_j(\cdot)$ using the methods described above, and then obtains an estimated $100(1-\alpha)$-quantile of $\sup_{\bx\in\mathcal{X}}|\widehat{Z}_j(\bx)|$ or $\sup_{\bx\in\mathcal{X}}|\widehat{z}^*_j(\bx)|$ given the data, denoted by $q_j(1-\alpha)$. Then, $(1-\alpha)$ confidence band for $\partial^\bq\mu(\bx)$ is given by
\[\widehat{\partial^\bq\mu}_j(\bx)\pm q_j(1-\alpha)\sqrt{\widehat{\Omega}_j(\bx)/n}.
\]
For example, the following command requests a critical value for constructing confidence bands:
\begin{example*}
\input{output/lspartition_10.txt}
\end{example*}

Once the critical value is available, the command \code{lsprobust.plot()} is able to visualize confidence bands:
\begin{example*}
\input{output/lspartition_11.txt}
\end{example*}
\vspace{-.2in}
\begin{figure}[!ht]
	\centering
	\includegraphics[scale=0.5]{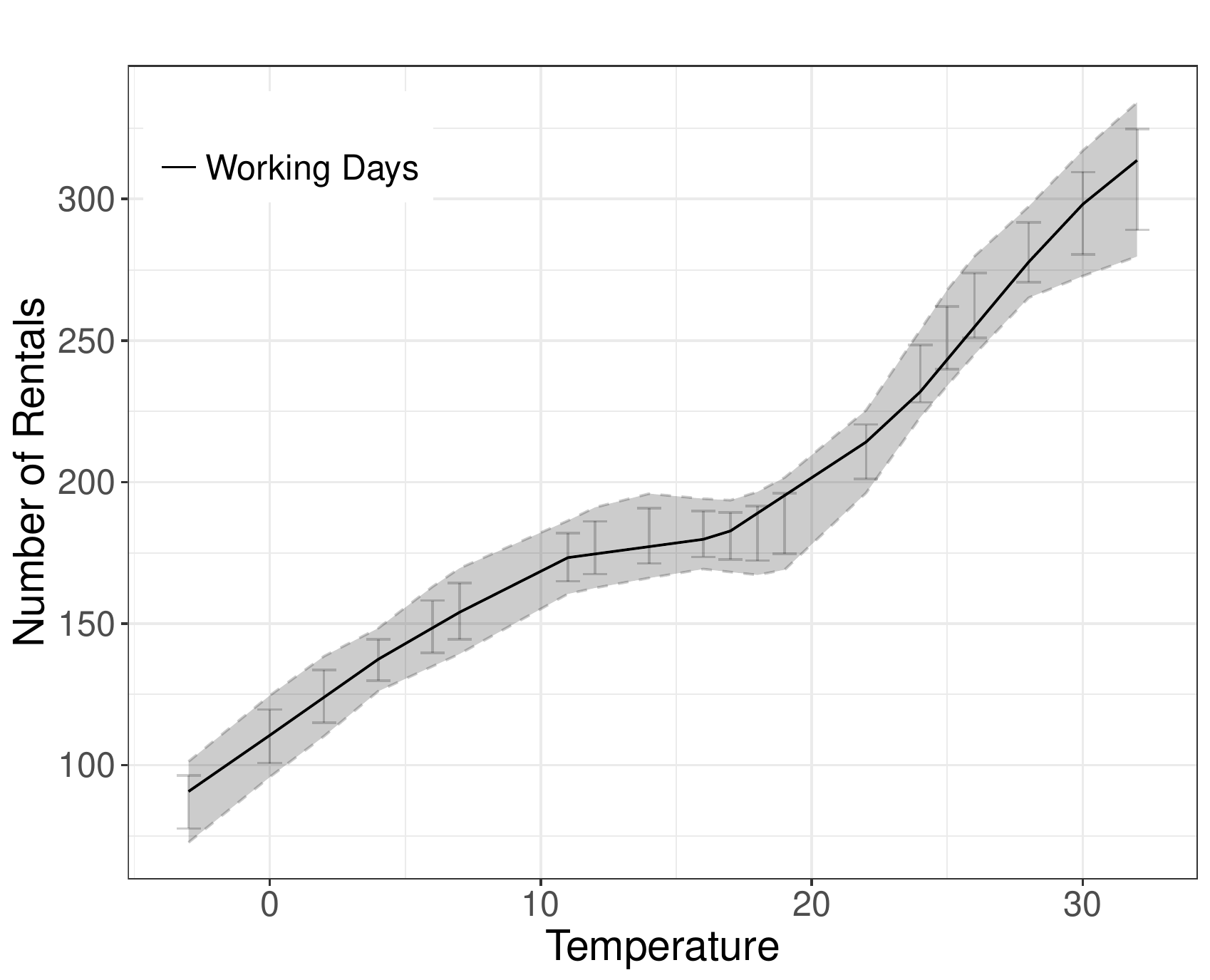}
	\caption{Robust inference: plug-in method, higher-order-basis basis correction}
	\label{figure:cb plug-in}
\end{figure}
The result is displayed in Figure \ref{figure:cb plug-in}. Since we set \code{CS="all"}, the command simultaneously plots pointwise confidence intervals (error bars) and a uniform confidence band (shaded region).

It is also possible to specify other bias correction approaches or uniform methods:
\begin{example*}
\input{output/lspartition_12.txt}
\end{example*}
\vspace{-.1in}

\begin{figure}[!ht]
	\centering
	\includegraphics[scale=0.5]{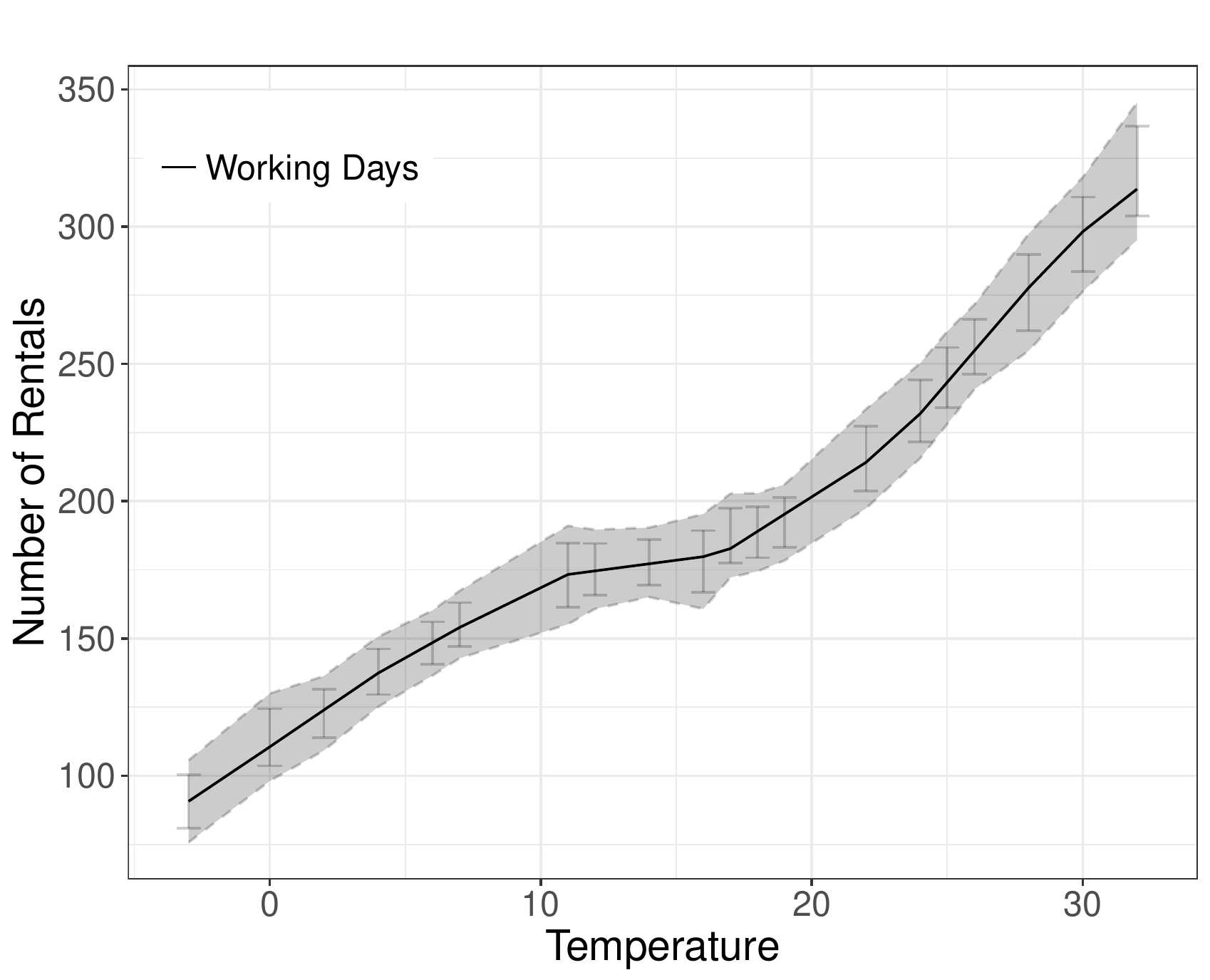}
	\caption{Robust inference: bootstrap method, plug-in bias correction}
	\label{figure:cb bootstrap}
\end{figure}
The result is displayed in Figure \ref{figure:cb bootstrap}. In this example, the critical values based on different methods are quite close, but in general their difference could be more pronounced in finite samples. See \citetalias{Cattaneo-Farrell-Feng_2019_AoS} for some simulation evidence.

\subsection{Linear combinations}
The package \pkg{lspartition} also includes a function \code{lsplincom()}, which implements estimation and inference for a linear combination of regression functions of different subgroups. To be concrete, consider a random trial with $G$ groups. Let $\mu(\bx; g)$  be the conditional expectation function (CEF) for group $g$, $g=1, \ldots, G$. The parameter of interest is $\theta(\bx)=\sum_{g=1}^{G}r_g\partial^\bq\mu(\bx; g)$, i.e., a linear combination of CEFs (or derivatives thereof) for different groups. To fix ideas, consider the most common application, the difference between two groups (or the conditional average treatment effect). Here, $G=2$, $\bq=\mathbf{0}$, and $(r_1, r_2)=(-1, 1)$. Then $\theta(\bx)  = \mathbb{E}[y_i|\bx_i=\bx, g=1] - \mathbb{E}[y_i|\bx_i=\bx, g=0]$.

To implement estimation and inference for $\theta(\bx)$, \code{lsplincom()} first calls \code{lsprobust()} to obtain a point estimate $\widehat{\partial^\bq\mu}_0(\bx;g)$ and all other objects for each group. The tuning parameter for each group can be selected by the data-driven procedures above. Then the point estimate of $\theta(\bx)$ is
\[\widehat{\theta}_0(\bx)=\sum_{g=1}^{G}r_g\widehat{\partial^\bq\mu}_0(\bx).
\]

The standard error of $\widehat{\theta}_j(\bx)$ can be obtained simply by taking the appropriate linear combination of standard errors for each $\widehat{\partial^\bq\mu}_j(\bx; g)$ and their estimated covariances. Robust confidence intervals can be similarly constructed as in \eqref{eq: CI}.

\code{lsplincom()} also allows users to construct confidence bands for $\theta(\cdot)$. Specifically, it requests \code{lsprobust()} to output the numerators (\code{t.num.pl} for ``plug-in'', or \code{t.num.wb1} and \code{t.num.wb2} for ``bootstrap'') and denominators (\code{t.denom}) of the feasible approximation processes $\widehat{Z}_j(\cdot)$ or $\widehat{z}^*(\cdot)$. Let $\bU_j(\cdot;g)$ and $\bv_j(\cdot;g)$ denote the numerator and denominator from group $g$ based on bias correction approach $j$, $g=1, \ldots, G$ and $j=1,2,3$. The approximation process for the $t$-statistic process based on $\widehat{\theta}_j(\bx)$ is
\[\widehat{Z}_{j, \theta}(\cdot)=
\frac{\sum_{g=1}^{G} r_g \bU_j(\cdot; g) \mathsf{N}_{g, K_{j,g}}}
{\sqrt{\sum_{g=1}^{G}r_g^2\bv_{j,g}(\cdot)^2}}
\]
where $\{\mathsf{N}_{g, K_{j,g}}\}_{g=1}^G$ is a collection of independent standard normal vectors, and $K_{j,g}$ indicates the dimension of $\mathsf{N}_{g, K_{j,g}}$. As discussed before, the dimensionality of these normal vectors depends on the particular bias correction approach and may vary across groups since the selected number of knots may be different across groups. The bootstrap approximation process $\widehat{z}^*_{j, \theta}(\cdot)$ can be constructed similarly.

Given these processes, inference is implemented by sampling from $G$ standard normal vectors (``plug-in" method) or $G$ groups of Rademacher vectors given the data. Then critical values used to construct $100(1-\alpha)$ confidence bands for $\theta(\cdot)$ are estimated similarly by $100(1-\alpha)$ empirical quantiles of $\sup_{\bx\in\mathcal{X}}|\widehat{Z}_{j,\theta}(\bx)|$ or $\sup_{\bx\in\mathcal{X}}|\widehat{z}^*_{j, \theta}(\bx)|$.

As an illustration, we compare the number of rentals during working days and other time periods (weekends and holidays) based on linear splines and plug-in bias correction. To begin with, we first estimate the conditional mean function for each group using the command \code{lsprobust()}.
\begin{example*}
\input{output/lspartition_13.txt}
\end{example*}
\vspace{-.1in}

\begin{figure}[!ht]
	\centering
	\includegraphics[scale=0.5]{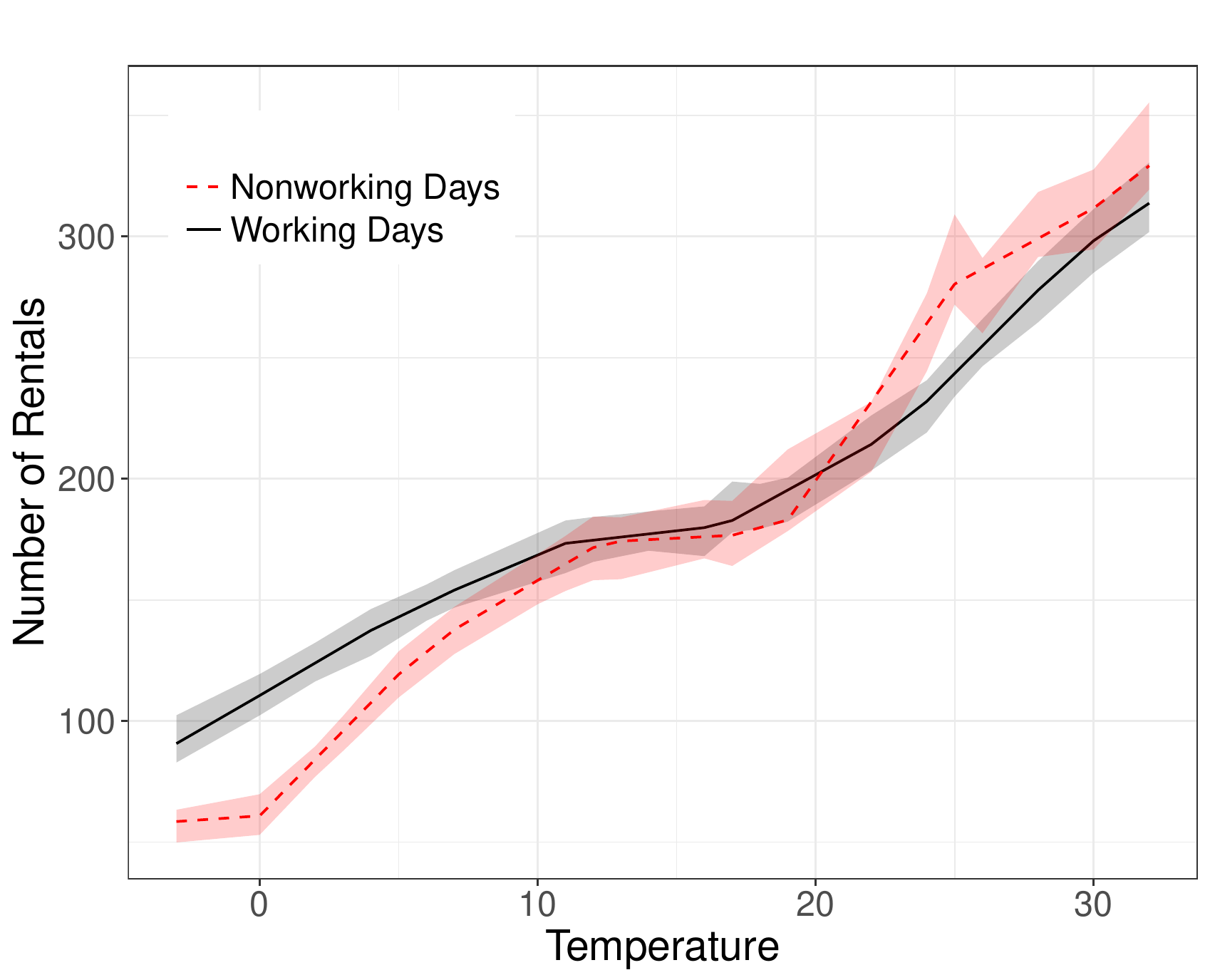}
	\caption{Point estimation and robust confidence intervals for two groups}
	\label{figure:diff1}
\end{figure}

The pointwise results for each group are displayed in Figure \ref{figure:diff1}. The shaded regions represent confidence intervals. Clearly, when the temperature is low, two regions are well separated, implying that people may rent bikes more during working days than weekends or holidays when the weather is cold.

Next, we employ the command \code{lsplincom()} to formally test this result. We specify \code{R=(-1, 1)}, denoting that $-1$ is the coefficient of the conditional mean function for the group \code{workingday==0} and $1$ is the coefficient of the conditional mean function for the group \code{workingday==1}.
\begin{example*}
\input{output/lspartition_14.txt}
\end{example*}
\vspace{-.1in}

The pointwise results are summarized in the above table. Clearly, when the temperature is low, the point estimate of the rental difference is significantly positive since the robust confidence intervals do not cover $0$. In contrast, when the temperature is above 7, it is no longer significant. This implies that the difference in the number of rentals between working days and other periods is less pronounced when the weather is warm. Again, we can use the command \code{lsprobust.plot()} to plot point estimates, confidence intervals and uniform band simultaneously:

\begin{example*}
\input{output/lspartition_15.txt}
\end{example*}
\vspace{-.2in}

In addition, some basic options for the command \code{lsprobust()} may be passed on to the command \code{lsplincom()}. For example, the following code generates a smoother fit of the rental difference by setting \code{m=3}:

\begin{example*}
	\input{output/lspartition_16.txt}
\end{example*}
\vspace{-.2in}

\begin{figure}[!ht]
	\centering
	\subfloat[$m=2$ \label{figure:m=2}]{\includegraphics[width=0.47\textwidth]{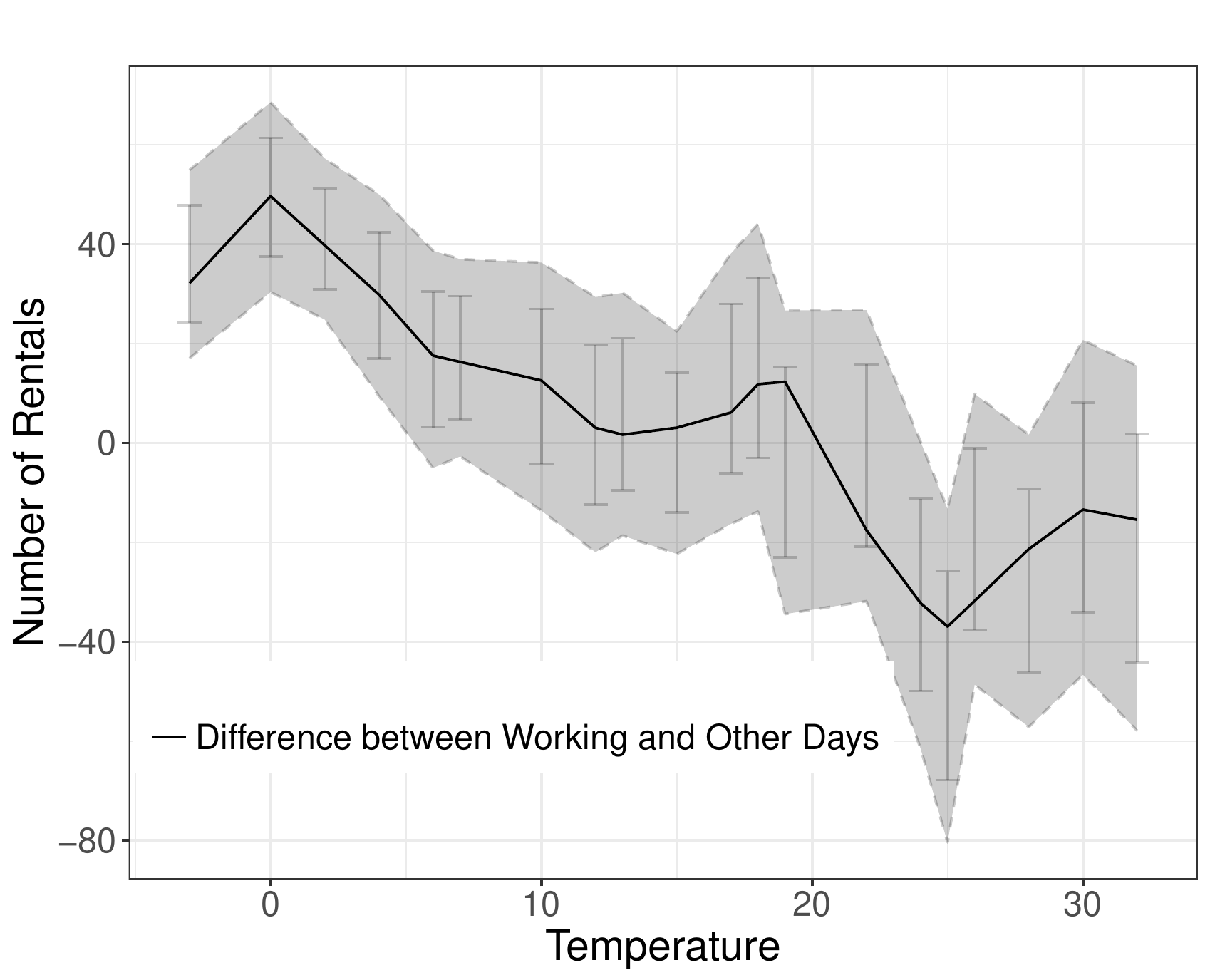}}
	~
	\subfloat[$m=3$ \label{figure: m=3}]{\includegraphics[width=0.47\textwidth]{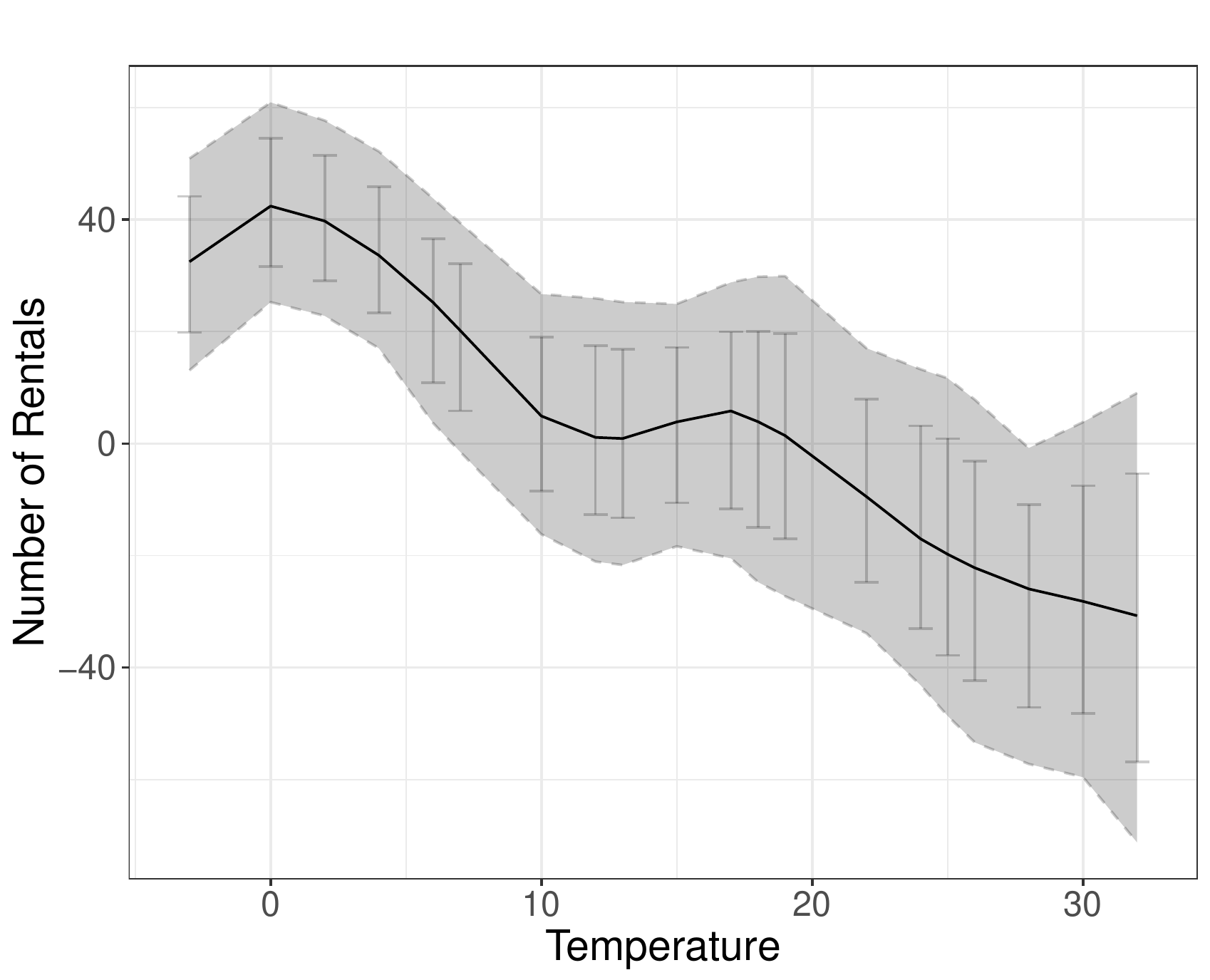}}
	\caption{Point estimation and robust inference: rental difference}
	\label{figure:diff}
\end{figure}

The results are shown in Figure \ref{figure:diff}. The confidence band for the difference is constructed based on the plug-in distributional approximation computed previously. It leads to an even stronger conclusion: the entire difference as a function of temperature is significantly positive \textit{uniformly} over a range of low temperatures since the confidence band is above zero when the temperature is low.

\section{Summary}

We gave an introduction to the software package \pkg{lspartition}, which offers estimation and robust inference procedures (both pointwise and uniform) for partitioning-based least squares regression. In particular, splines, wavelets, and piecewise polynomials are implemented. The main underlying methodologies were illustrated empirically using real data. Finally, installation details, scripts replicating the numerical results reported herein, links to software repositories, and other companion information, can be found in the package's website:

\begin{center}\url{https://sites.google.com/site/lspartition/}.\end{center}

\bibliography{Cattaneo-Farrell-Feng}

\begin{thebibliography}{13}
\providecommand{\natexlab}[1]{#1}
\providecommand{\url}[1]{\texttt{#1}}
\expandafter\ifx\csname urlstyle\endcsname\relax
  \providecommand{\doi}[1]{doi: #1}\else
  \providecommand{\doi}{doi: \begingroup \urlstyle{rm}\Url}\fi

\bibitem[Calonico et~al.(2018)Calonico, Cattaneo, and
  Farrell]{Calonico-Cattaneo-Farrell_2018_JASA}
S.~Calonico, M.~D. Cattaneo, and M.~H. Farrell.
\newblock On the effect of bias estimation on coverage accuracy in
  nonparametric inference.
\newblock \emph{Journal of the American Statistical Association}, 113\penalty0
  (522):\penalty0 767--779, 2018.

\bibitem[Calonico et~al.(2019)Calonico, Cattaneo, and
  Farrell]{Calonico-Cattaneo-Farrell_2019_wp}
S.~Calonico, M.~D. Cattaneo, and M.~H. Farrell.
\newblock Coverage error optimal confidence intervals for local polynomial
  regression.
\newblock \emph{\emph{arXiv:1808.01398}}, 2019.

\bibitem[Cattaneo and Farrell(2013)]{Cattaneo-Farrell_2013_JoE}
M.~D. Cattaneo and M.~H. Farrell.
\newblock Optimal convergence rates, bahadur representation, and asymptotic
  normality of partitioning estimators.
\newblock \emph{Journal of Econometrics}, 174\penalty0 (2):\penalty0 127--143,
  2013.

\bibitem[Cattaneo et~al.(2019)Cattaneo, Farrell, and
  Feng]{Cattaneo-Farrell-Feng_2019_AoS}
M.~D. Cattaneo, M.~H. Farrell, and Y.~Feng.
\newblock Large sample properties of partitioning-based estimators.
\newblock \emph{Annals of Statistics}, page forthcoming, 2019.

\bibitem[Chui(2016)]{Chui_2016_Book}
C.~K. Chui.
\newblock \emph{An Introduction to Wavelets}.
\newblock Elsevier, 2016.

\bibitem[Cohen et~al.(1993)Cohen, Daubechies, and Vial]{Cohen_1993_ACHA}
A.~Cohen, I.~Daubechies, and P.~Vial.
\newblock Wavelets on the interval and fast wavelet transforms.
\newblock \emph{Applied and Computational Harmonic Analysis}, 1\penalty0
  (1):\penalty0 54--81, 1993.

\bibitem[Fan and Gijbels(1996)]{Fan-Gijbels_1996_Book}
J.~Fan and I.~Gijbels.
\newblock \emph{Local Polynomial Modelling and Its Applications}.
\newblock Chapman \& Hall/CRC, New York, 1996.

\bibitem[Gy{\"o}rfi et~al.(2002)Gy{\"o}rfi, Kohler, Krzy{\.z}ak, and
  Walk]{Gyorfi-etal_2002_book}
L.~Gy{\"o}rfi, M.~Kohler, A.~Krzy{\.z}ak, and H.~Walk.
\newblock \emph{A Distribution-Free Theory of Nonparametric Regression}.
\newblock Springer-Verlag, 2002.

\bibitem[Harezlak et~al.(2018)Harezlak, Ruppert, and
  Wand]{Harezlak-Ruppert-Wand_2018_book}
J.~Harezlak, D.~Ruppert, and M.~P. Wand.
\newblock \emph{Semiparametric Regression with R}.
\newblock Springer, New York, 2018.

\bibitem[Long and Ervin(2000)]{Long-Ervin_2000_AS}
J.~S. Long and L.~H. Ervin.
\newblock Using heteroscedasticity consistent standard errors in the linear
  regression model.
\newblock \emph{The American Statistician}, 54\penalty0 (3):\penalty0 217--224,
  2000.

\bibitem[Ruppert et~al.(2009)Ruppert, Wand, and
  Carroll]{Ruppert-Wand-Carroll_2009_book}
D.~Ruppert, M.~P. Wand, and R.~Carroll.
\newblock \emph{Semiparametric Regression}.
\newblock Cambridge University Press, New York, 2009.

\bibitem[Wickham and Chang(2016)]{ggplot2}
H.~Wickham and W.~Chang.
\newblock \emph{ggplot2: Create Elegant Data Visualisations Using the Grammar
  of Graphics}, 2016.
\newblock URL \url{https://CRAN.R-project.org/package=ggplot2}.
\newblock R package version 2.2.1.

\bibitem[Zhang and Singer(2010)]{Zhang-Singer_2010_Book}
H.~Zhang and B.~H. Singer.
\newblock \emph{Recursive Partitioning and Applications}.
\newblock Springer, 2010.

\end{thebibliography}

\address{Matias D. Cattaneo\\
	Department of Operation Research and Financial Engineering\\
	Princeton University\\
	Princeton, NJ 08544\\
	\email{cattaneo@princeton.edu}}

\address{Max H. Farrell\\
	Booth School of Business\\
	University of Chicago\\
	Chicago, IL 60637\\
	\email{max.farrell@chicagobooth.edu}}

\address{Yingjie Feng\\
	Department of Politics\\
	Princeton University\\
	Princeton, NJ 08544\\
	\email{yingjief@princeton.edu}}

\end{article}

\end{document}